\documentclass[a4paper,fleqn,usenatbib]{mnras}

\usepackage[T1]{fontenc}
\usepackage{ae,aecompl}

\usepackage{graphicx}	
\usepackage{amsmath}	
\usepackage{amssymb}	
\usepackage{pdflscape}
\usepackage{multirow}
\usepackage{subfig}

\title[A cosmology-independent calibration of SNe Ia data]{A cosmology-independent calibration of type Ia supernovae data}

\author[C. Hauret et al.]{
C. Hauret$^{1}$\thanks{E-mail:clementine.hauret@ulg.ac.be },
P. Magain$^{1}$ 
and J. Biernaux$^{1}$\\
$^{1}$STAR-OrCA, Universit\'e de Li\`ege, All\'ee du 6 Ao\^ut 19c, B-4000 Li\`ege, Belgium
}

\date{Accepted XXX. Received YYY; in original form ZZZ}

\pubyear{2018}

\begin{document}
\label{firstpage}
\pagerange{\pageref{firstpage}--\pageref{lastpage}}
\maketitle

\begin{abstract}
Recently, the common methodology used to transform type Ia supernovae (SNe Ia) into genuine standard candles has been suffering criticism. Indeed, it assumes a particular cosmological model (namely the flat $\Lambda$CDM) to calibrate the standardisation corrections parameters, i.e. the dependency of the supernova peak absolute magnitude on its colour, post-maximum decline rate and host galaxy mass. As a result, this assumption could make the data compliant to the assumed cosmology and thus nullify all works previously conducted on model comparison. In this work, we verify the viability of these hypotheses by developing a cosmology-independent approach to standardise SNe Ia data from the recent JLA compilation. Our resulting corrections turn out to be very close to the $\Lambda$CDM-based corrections. Therefore, even if a $\Lambda$CDM-based calibration is questionable from a theoretical point of view, the potential compliance of SNe Ia data does not happen in practice for the JLA compilation. Previous works of model comparison based on these data do not have to be called into question. However, as this cosmology-independent standardisation method has the same degree of complexity than the model-dependent one, it is worth using it in future works, especially if smaller samples are considered, such as the superluminous type Ic supernovae.
\end{abstract}




\section{Introduction}

Since the end of the 1990s, modern cosmology has been developed around the idea that the most accurate representation of the contents and evolution of our universe is the flat $\Lambda$CDM model, a model of a spatially flat universe containing baryonic and dark matter as well as a repulsive component called dark energy. Over the years, this model proved to be extremely predictive and consistent with the many and varied observations of type Ia supernovae (SNe Ia), cosmological microwave background or large-scale structures, to name but a few. Nowadays, this model is so well-established that it has become the standard model of cosmology, totally outshining any other cosmological model. 

At the origin of the 20-year reign of this model, we find the observations of SNe Ia \citep{Riess98, Perlmutter99}. These objects are extremely important for cosmologists as they are nearly perfect standard candles, observable over large distances. An astrophysical object is a standard candle if its intrinsic luminosity is known or can be determined independently of its distance. One can thus deduce the distance of such an object by measuring its apparent flux. 

SNe Ia are believed to arise from thermonuclear explosions of white dwarfs in a binary system \citep{Hoyle60}. As these white dwarfs accrete matter from their companions, they grow and reach explosion conditions when their mass approaches the Chandrasekhar limit \citep{Chandra31}. If the very details of this scenario are still a matter of debate, it is reasonable to assume that all SNe Ia  present roughly the same luminosity as the explosions of white dwarfs of similar masses should produce the same amount of energy. Therefore, these objects make good standard candles.

However, in the last decades, small but significant variations of their peak luminosities have been observed, implying that some corrections have to be applied in order to use SNe Ia as genuine standard candles. Correlations have  been found between the intrinsic brightness of SNe Ia, the shape of their light-curve \citep{Phillips93}, their colour \citep{Tripp98} and their host galaxy \citep{Kelly10, Sullivan10, Lampeitl10}. Thanks to these correlations, SNe Ia are nowadays considered as one of the best cosmological tools.

The methodology to deal with these light-curve shape, colour and host galaxy corrections (called hereafter the standardisation corrections) is extremely important in order to accurately describe our universe via cosmological models. Over the years, a particular approach appeared \citep{Perlmutter99} and developed amongst the community. Nowadays widely approved and used \citep[e.g.][for recent examples]{Suzuki12,Betoule14,Calcino17}, this methodology is unfortunately not perfect. Indeed, it assumes a cosmological model, having the potential collateral effect of making the data compliant to this particular model, disfavouring any other cosmologies \citep{Vish10, Marriner11, Melia12, Wei15}. Corrected that way, the SNe Ia could no longer be used to compare different models of the universe and works to determine the most representative one (such as the ones recently published by \cite{Xu16} or \cite{Hee17}) could then be void and obsolete.

In this paper, we develop a simple and completely cosmology-independent approach to standardise SNe Ia data and to check whether this problem actually occurs in the JLA compilation data\footnote{\url{http://supernovae.in2p3.fr/sdss_snls_jla/ReadMe.html}} \citep{Betoule14}. Hence in Sec. \ref{Current}, we explain in further details the currently used methodology to process SNe Ia standardisation corrections and we highlight its potential issues. Then, we present our alternative calibration and answer the question of the possible compliance of data in Sec. \ref{Alt}. 

\section{The current methodology to process SNe Ia data}
\label{Current}

In order to be considered as genuine standard candles, the SNe Ia peak luminosities have to be corrected from the standardisation corrections. Mathematically speaking, following the works of \citet{Phillips93}, \citet{Tripp98} and \citet{Suzuki12}, we can compute the corrected peak absolute magnitude as:
\begin{equation}
M_{B, \mathrm{corr}} = M_{B} - \alpha x_1 + \beta c + \delta P\left( M_{\mathrm{stellar}} < 10^{10} M_{\sun} \right).
\label{eq:lumcorr}
\end{equation}
$M_{B}$ is the absolute blue magnitude of a type Ia supernova (SN Ia) with $x_1$, $c$ and $P=0$ while $M_{B, \mathrm{corr}}$ is this magnitude for a SN Ia of given $x_1$, $c$ and $P$. The latter are parameters linked respectively to the SN Ia light-curve shape, its  colour and its host galaxy mass (see below), while $\alpha$, $\beta$ and $\delta$ are parameters that describe the correlations of the peak magnitude to the three aforementioned properties. 

First, $x_1$ is a measurement of the light-curve shape, related to the so-called {\em stretch correction} as the differences in light-curve shape can also be seen as the stretching of the light-curve time axis \citep{Perlmutter97,Perlmutter97ip}. Second, $c$ is a parameter characterising the colour of the object \citep{Tripp98}. Finally, $P$ is the probability that the SN Ia host galaxy is less massive than a threshold fixed at $10^{10} M_{\sun}$ \citep{Sullivan10}. One will notice that the relation between host mass and absolute magnitude presented in Eq. \ref{eq:lumcorr} can be implemented differently than with the probability $P$ defined in this article or in others such as \cite{Suzuki12} or \cite{Mohlabeng14}. For example, it can be computed as a step function of the host mass as it has been done in \cite{Betoule14} or in the very recent \cite{Calcino17}.

Initially, $M_{B}$ as well as the $\alpha$ and $\beta$ coefficients\footnote{One can notice that the $\delta$ coefficient was not yet introduced at that time but it can nevertheless be calibrated the same way.} were calibrated on nearby SNe Ia \citep[e.g.][as well as papers based on the MLCS light-curve fitter introduced by \citet{Riess96}]{Phillips93, Hamuy95, Tripp97, Tripp98}  and the relations were applied to more distant objects. However, since \citet{Perlmutter99}, another determination of these standardisation corrections has been introduced and developed around the SALT and SALT2 light-curve fitters \citep{Guy05, Guy07}. Nowadays, this methodology is the most common way to transform SNe Ia into standard candles. $M_{B}$, $\alpha$, $\beta$, and $\delta$ are seen as nuisance parameters and are determined together with the cosmological parameters by fitting an assumed model (generally the flat $\Lambda$CDM model) on the Hubble diagram \citep{Suzuki12, Betoule14, Calcino17}. 

Unfortunately, while it gives better fits if the flat $\Lambda$CDM model is in fact the best representation of our universe, this technique can lead to a bias in its favour \citep{Vish10, Marriner11, Melia12, Wei15}. Indeed, when the cosmology and the standardisation corrections are fitted simultaneously, their respective parameters are no longer independent from each other. In theory, this could skew the observational data towards the assumed cosmology (a flat $\Lambda$CDM model in the case we are interested in), at the expense of any other one. This will especially be true if the standardisation corrections are correlated with the redshift. With such corrections, the SNe Ia data could no longer be used to test cosmological models as they would always favour the assumed one. In order to correctly compare the ability of various cosmological models to fit the SNe Ia data, the standardisation parameters should be redetermined for each non-$\Lambda$CDM cosmologies (as it has been done by \cite{Wei15} but not in model comparison works we are interested in such as \cite{Xu16} or \cite{Hee17}), leading to possible time-consuming calculations. 

Of course, nowadays the flat $\Lambda$CDM model is widely favoured by various observations. Hence, most of the present observational works, including SNe Ia analyses, are not aimed any more at testing which cosmological model is the most representative of our universe but they are instead designed to increase the precision on the determination of its density parameter $\Omega_{\mathrm{m},0}$. In this case, the methodology currently used for SNe Ia is perfectly well-suited. However, its possible collateral effect on data and on model comparison raises important philosophical questions. Should we approve of a way of processing the data which can potentially make them compliant to a certain theoretical model? Should we not prefer a cosmology-independent approach to avoid this possible problematic effect?

\section{Developing a cosmology-independent calibration of SNe Ia data}
\label{Alt}

To answer these questions and check if this potential compliance of data is effectively present in the SNe Ia compilations, we develop an alternative cosmology-independent calibration to process the standardisation corrections. Different cosmological models could then be fitted on the recalibrated SNe Ia data in order to determine the most representative model of our universe as it has been done by \cite{Xu16} for example. The decisive advantage of the standardisation parameters values we determine is that they can be directly used when working with non-$\Lambda$CDM cosmologies instead of having to be re-evaluated for each studied cosmological model.

\subsection{Calibration principles}
\label{Alt-principles}

Apart from developing a precise model able to determine the standardisation parameters directly from the physics of the SNe Ia thermonuclear explosions, the only way to evaluate these parameters is through the Hubble diagram. Indeed, on the one hand, one can express the distance modulus $\mu$ of each SN Ia as a difference between its apparent and absolute magnitude (respectively $m_B$ and $M_{B, \mathrm{corr}}$) and, using Eq. \ref{eq:lumcorr}, we have 
\begin{equation}
\mu = m_B - M_B  + \alpha x_1 - \beta c - \delta P\left( M_{\mathrm{stellar}} < 10^{10} M_{\sun} \right).
\label{eq:mu_lum}
\end{equation}

On the other hand, this distance modulus can also be found from the SN Ia redshift $z$ via the luminosity distance $d_\mathrm{L}(z)$:
\begin{equation}
\mu = 5 \log \left(d_\mathrm{L}(z) \right) - 5.
\label{eq:mu_dl}
\end{equation}
Unfortunately, the $d_\mathrm{L}(z)$ relation cannot be found exactly without assuming a particular cosmological model. Hence, as previously said, the recent SNe Ia compilations have been primarily processed assuming a flat $\Lambda$CDM model but one can use other cosmology, such as the $R_\mathrm{h}=ct$ model \citep{Wei15} for example. However, this prior choice of a model could lead to the already-pointed-out problem of data compliance. 

To put this possibility to the test, we decide to get rid of any assumption on the cosmology at the cost of loosing the exact dependency of luminosity distance with redshift. To do so, we expand the $d_\mathrm{L}(z)$ relation as a Taylor series around the observer (for whom $z=0$). We test two different Taylor expansions: (i) one of the redshift (hereafter called the $z$-series):
\begin{equation}
d_\mathrm{L}(z) = d_{0,z} ~\left( z + d_{1,z} z^2 + d_{2,z} z^3 + d_{3,z} z^4 + \mathrm{O}(z^5) \right) 
\label{eq:dl_taylor_z}
\end{equation}
and (ii) one of a function $f$ of the redshift ($f(z) = z/ 1+z$) (hereafter called the $f$-series):
\begin{multline}
d_\mathrm{L}(z) = d_{0,f} ~\Big( f(z) + d_{1,f} f^2(z) \\ + d_{2,f} f^3(z)   + d_{3,f} f^4(z) + \mathrm{O}\left(f^5(z)\right) \Big). 
\label{eq:dl_taylor_f}
\end{multline}
One can directly notice that in both cases, we logically impose a zero luminosity distance at a zero redshift. In Subsec. \ref{Alt-calib}, we will choose between these two kinds of series, limited to different orders, so as to minimise the number of Taylor parameters while optimising the representation of SNe Ia data.

Our specific choice of the $f(z)$ function can be explained as follows. As SNe Ia with redshifts larger than unity have been observed, the high-order terms of the $z$-series could quickly become non-negligible if their associated parameters $\{d_{i,z}\}$ are non-null. The $d_\mathrm{L}(z)$ relation would then not be well approximated by a limited series and our calibration would be useless. This is why we also consider the $f$-series with the $f$ function such that it stays low as $z$ goes high. We would then \textit{a priori} neglect high-order terms more easily than in the $z$-series. Furthermore, one can notice that in a model based on the Friedmann-Lema\^itre-Robertson-Walker (FLRW) metric, this $f$-series is in fact the luminosity distance expanded as a Taylor series of the scale factor $R$ of the universe around its actual value $R(t_0)$ as $z+1 = R(t_0)/R$. Our particular choice of the function $f(z)$ is thus quite natural in a FLRW context.

To sum up, with these two Taylor series, all the information about cosmology is exclusively included in the Taylor parameters $\{d_{i, z/f}\}$. We would like to point out that in neither of our series, do we suppose a specific form of the metric. We do not even assume General Relativity. Nonetheless, for cosmological models based on General Relativity and on the FLRW metric, these parameters are directly related to the well-known present Hubble $H_0$ and deceleration $q_0$ parameters defined by
\begin{equation*}
H_0 \equiv H(t_0) = \frac{\dot{R}(t_0)}{R(t_0)}
\mathrm{~~and~~ } q_0 \equiv q(t_0) = - \frac{\ddot{R}(t_0)}{R(t_0)}  \left[ \frac{\dot{R}(t_0)}{R(t_0)} \right]^{-2}.
\label{eq:h0q0j0}
\end{equation*}
Hereafter, we will set the scale factor present value $R(t_0)$ at unity. In this way, the first two Taylor parameters of the $z$-series have a very simple form:  
\begin{equation*}
d_{0,z} = \frac{c_\mathrm{light}}{H_0} \mathrm{~~and ~~}d_{1,z} = \frac{1-q_0}{2} 
\label{eq:d0d1d2}
\end{equation*}
where $c_\mathrm{light}$ is the speed of light. For completeness, the expressions of the next two Taylor parameters of this same series can be found in \citet{Visser04}.

Going back to our estimate of the standardisation parameters through the Hubble diagram, we can combine Eqs. \ref{eq:mu_lum}, \ref{eq:mu_dl} and \ref{eq:dl_taylor_z} or \ref{eq:dl_taylor_f} (limiting here the series to the third order) to have
\begin{multline}
m_B - M_B + \alpha x_1 - \beta c - \delta P\left( M_{\mathrm{stellar}} < 10^{10} M_{\sun} \right) \\ \simeq 5 \log(d_{0,z}) + 5 \log\left( z + d_{1,z} z^2 + d_{2,z} z^3 \right)  - 5
\label{eq:mod_ourcal_z}
\end{multline}
or
\begin{multline}
m_B - M_B + \alpha x_1 - \beta c - \delta P\left( M_{\mathrm{stellar}} < 10^{10} M_{\sun} \right)  \\ \simeq 5 \log(d_{0,f}) + 5 \log\left( f(z) + d_{1,f} f^2(z) + d_{2,f} f^3(z) \right) - 5 .
\label{eq:mod_ourcal_f}
\end{multline}
The terms  appearing in these equations are divided in two categories: (i) the SN Ia observable properties ($m_B, x_1, c, P$ and $z$) and (ii) the standardisation and Taylor parameters ($\alpha$, $\beta$, $\delta$, $d_{i\ne0, z/f}$ as well as a combination\footnote{For models based on the FLRW metric, we find the well-known degeneracy between $M_B$ and $H_0$.} of $M_B$ and $d_{0, z/f} $). The latter can be determined by minimising a $\chi^2$ statistics (as it has been made in \cite{Suzuki12} and \cite{Betoule14}) or, in a more rigorous approach, by using a bayesian framework (such as the one first developed by \cite{March11} and its extensions applied to the most recent SNe Ia compilation by \cite{Shariff16}, for example).

\subsection{Dataset and intrinsic dispersion}
\label{Alt-data}

In this work, we use SNe Ia data released in the JLA compilation from \cite{Betoule14}, containing 740 objects with redshifts ranging from around 0.01 to 1.3. 

When working with SNe Ia data, one has to evaluate a parameter besides the standardisation parameters: the intrinsic dispersion of the SNe Ia magnitudes. It may include both a true intrinsic scattering component and any uncharacterised sources of errors. 

To determine its value, we cannot use the usual methodology \citep[for examples]{Conley11, Sullivan11, Suzuki12}, which consists in adding an error  to that of the SNe Ia apparent magnitude until the best-fitting $\chi^2$ reaches a value of one per degree of freedom. This procedure obviously forbids any comparison between models as all fits end up with equally good (artificial) $\chi^2$ values. This problem is not new and has been pointed out by many authors such as \citet{Conley11}. Nevertheless, it is particularly an issue for our work as we eventually have to choose between the different Taylor series presented in the previous subsection.

Fortunately, \citet{Betoule14} developed a technique to avoid this problem and to evaluate this dispersion more properly. Basically, they divide the SNe Ia into different redshift bins and evaluate the intrinsic dispersion in each bin via a restricted log-likelihood criterion. For more information, we strongly encourage the reader to directly turn to Section 5.5 of \citet{Betoule14}. 

Our resulting intrinsic dispersion can be found in Table \ref{tab:disp}. One will notice that similarly to \citet{Betoule14} and authors before them \citep{Conley11, Sullivan11}, we determine different dispersions for each sample of objects at our disposal (\textit{i.e.} the low-z, SDSS-II, SNLS and HST samples from the JLA compilation). Indeed, the dispersion we want to evaluate includes a true intrinsic component as well as a component coming from the remaining unexplained sources of errors, which could vary from one survey program to another. Then, the dispersion associated to each sample in Table \ref{tab:disp} is the weighted mean of the dispersions evaluated on the different redshift bins defined by \cite{Betoule14}. 

\begin{table}
\caption{Magnitude intrinsic dispersion evaluated separately on each SNe Ia sample through the methodology developed by \citet{Betoule14}. Our values are compatible with previous determinations \citep{Conley11, Betoule14}.  }
\label{tab:disp}
\centering
\begin{tabular}{ c c }
\hline 
\hline
Sample & Intrinsic dispersion \\
\hline
low-$z$ & $0.116$  \\
SDSS-II & $0.045$  \\
SNLS & $0.128$ \\
HST & $0.060$  \\
\hline
\end{tabular}
\end{table}

\subsection{Cosmology-independent calibration}
\label{Alt-calib}

As stated in the two previous subsections, we have to choose between the $d_\textrm{L}$ $z$- and $f$-series (Eqs. \ref{eq:dl_taylor_z} or \ref{eq:dl_taylor_f}) and limit them to a particular order, so that we minimise the number of Taylor parameters while optimising the representation of our data. 

We calibrate the standardisation and Taylor parameters on SNe Ia data for these two series limited to five increasing orders. To determine the optimal of these 10 calibrations, we do not use the usual minimisation of a $\chi^2$ statistics as this criterion does not take into account the complexity (i.e. the number of parameters) of the series. Indeed, at equivalent agreement, we want to favour a model with fewer parameters. This has also the advantage to avoid an overtraining of our methodology. Other statistical criteria, called information criteria, have been developed to consider this idea of complexity. Among them, the Akaike information criterion \citep[AIC;][]{Akaike74} and the Bayesian information criterion \citep[BIC;][]{Schwarz78} are widely used. They are defined by
\begin{equation*}
\mathrm{AIC} = - 2 \ln \mathcal{L}_\mathrm{max} + 2 k  \mathrm{~~~~and~~~~}  \mathrm{BIC} = - 2 \ln \mathcal{L}_\mathrm{max} + k \ln N
\label{eq:aicbic}
\end{equation*}
where $\mathcal{L}_\mathrm{max}$ is the maximum likelihood ($- 2 \ln \mathcal{L}_\mathrm{max} = \chi^2_\mathrm{min}$ if we assume that the errors on the data are normally distributed, as it is generally the case in cosmology), $k$ is the number of parameters of the studied model and $N$ is the number of data points. 

The model with the lowest AIC and BIC will be preferred but only the relative values between different models are important and useful for comparison. Thus, to quantify the goodness of a given model $i$ compared to the one with the minimum AIC or BIC, one has to compute the differences $\Delta\mathrm{AIC}_i = \mathrm{AIC}_i - \mathrm{AIC}_\mathrm{min}$ or $\Delta\mathrm{BIC}_i = \mathrm{BIC}_i - \mathrm{BIC}_\mathrm{min}$. The larger these differences, the stronger the empirical evidence against the given model. 
More information about these criteria can be found in \cite{Liddle04} for example.

\begin{figure*}
\includegraphics[width=0.9\textwidth]{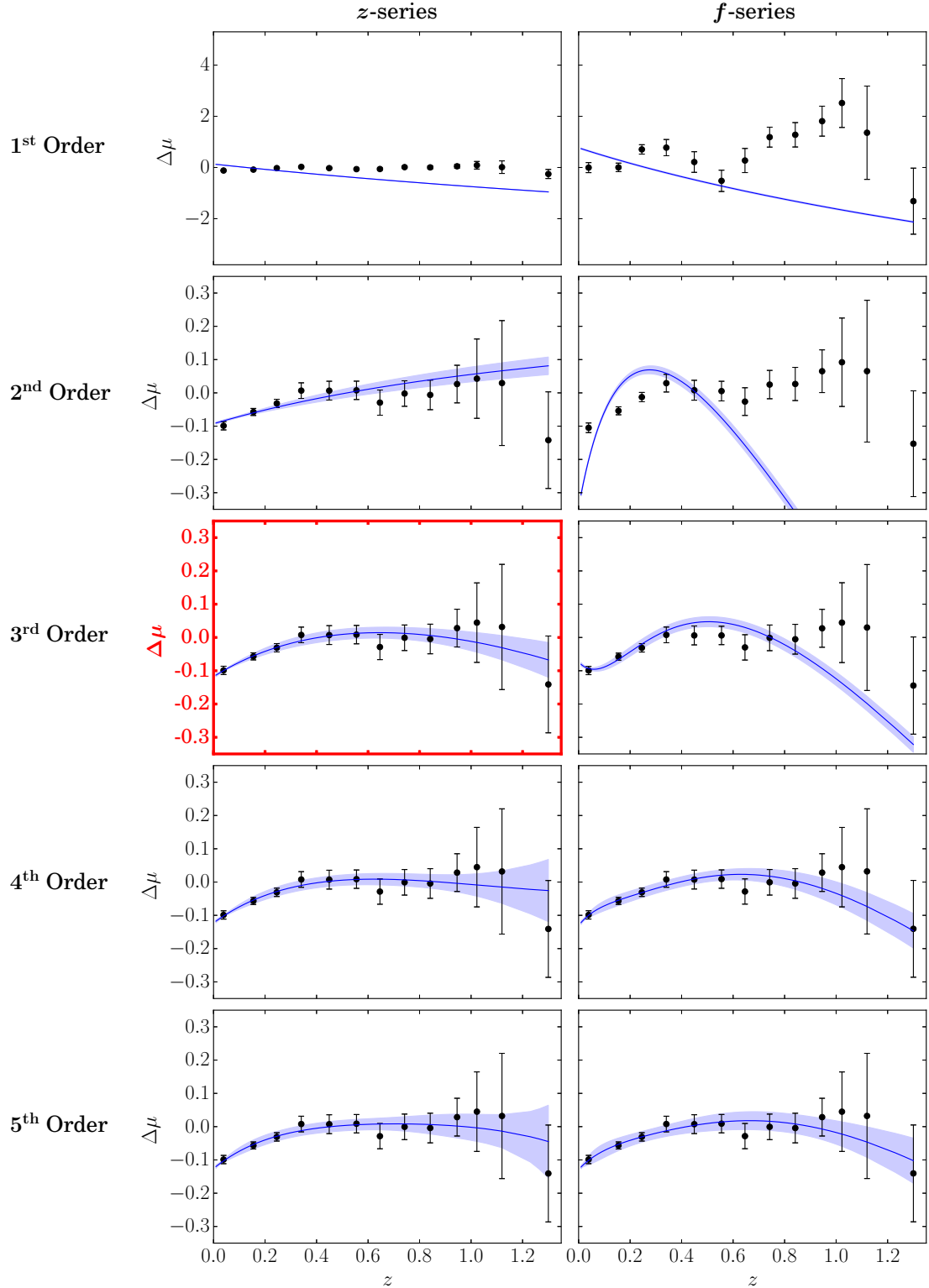}
\caption{Hubble diagram residuals when the distance moduli of an empty cosmological model have been subtracted. Each panel compares the weighted SNe Ia data over $0.1$-length redshift bins (built on the standardisation parameters presented in Table \ref{tab:cal-param-wodisp}; black points) with our calibration distance moduli (built on the Taylor parameters presented in this same Table; blue line with shaded zones corresponding to $1\sigma$ errors). \textit{From left to right}: calibration based on the $z$- or $f$-series. \textit{From top to bottom}: calibration based on the first to the fifth order of the series. Obviously, the $f$-series reproduce the behaviour of the data less correctly than the $z$-ones. Amongst the latter, the series limited to the third order (whose panel is highlighted in bold and red) minimises the number of Taylor parameters while optimising the representation of the data (see Table \ref{tab:cal-chi-wodisp}). }
\label{fig:hub_diag_wodisp}
\end{figure*}

Our resulting standardisation and Taylor parameters can be found in Table \ref{tab:cal-param-wodisp}. Each set of parameters is determined using a simulated annealing algorithm implementing a $\chi^2$ minimisation on the whole SNe Ia population, i.e. on the 740 objects. For each of these sets, the covariance matrix has been calculated, yielding the one-sigma error bars. Table \ref{tab:cal-chi-wodisp} displays the corresponding $\chi^2$ for these best-fitting models and their associated AIC and BIC, so that models can be compared to each other. These $d_\textrm{L}(z)$ series are also plotted on Hubble diagrams, compared to binned data, on Fig. \ref{fig:hub_diag_wodisp}.

First of all, it is clear in Table \ref{tab:cal-chi-wodisp} and on Fig. \ref{fig:hub_diag_wodisp} that the $z$-series represent the data more quickly and more correctly than the $f$-ones. Even the fifth order of the $f$-series does not provide a fit as good as the $z$-series, third order. This result is slightly surprising as one would \textit{a priori} be more confident in the $f$-series where the values $f(z) = z/1+z$ stays low while $z$ itself can reach relatively high values (there are 9 objects with $z\geqslant1$ in our dataset). Non-negligible differences between the true $d_\textrm{L}(z)$ relation and its Taylor approximation in $z$ could thus be expected at high redshift. Nevertheless, if the redshift powers of Eq. \ref{eq:dl_taylor_z} can be high for the most distant objects, these powers are multiplied by relatively low Taylor parameters $\{d_{i,z}\}$ as can be seen in Table \ref{tab:cal-param-wodisp}. This has to be compared with the lower powers of $f(z)$ (see Eq. \ref{eq:dl_taylor_f}), multiplied by the quite high associated $\{d_{i,f}\}$ values, presented in this same Table. 

\begin{figure}
\includegraphics[width=\columnwidth]{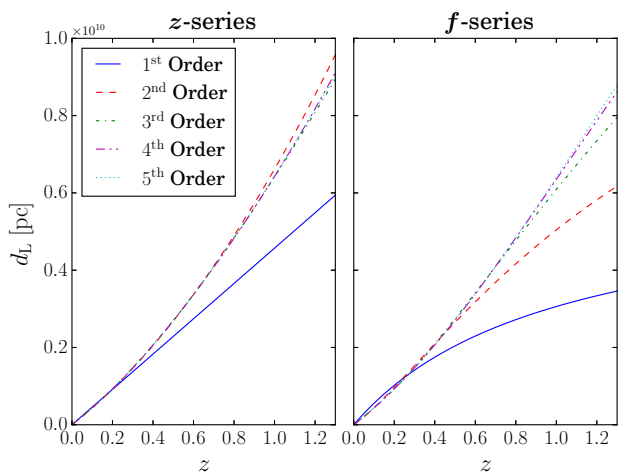}
\caption{Evolution of our best-fitting $z$- (left panel) and $f$-series (right panel) limited to different orders as a function of the redshift $z$. These approximated $d_\mathrm{L}(z)$ relations are given by Eqs. \ref{eq:dl_taylor_z} and \ref{eq:dl_taylor_f} whose Taylor parameters values $\{d_{i,z/f}\}$ can be found in Table \ref{tab:cal-param-wodisp}. The addition of supplementary terms in the $z$-series results in smaller modifications of the $d_\mathrm{L}(z)$ function than for the $f$-series. Despite an \textit{a priori} preference for the $f$-series to expand the $d_\mathrm{L}(z)$ function, the $z$ one eventually appears to be the one to use in this situation. }
\label{fig:dl(z)}
\end{figure}

This can also be visualised in Fig. \ref{fig:dl(z)} where the $z$ and $f$ Taylor series limited to different orders as a function of the redshift are plotted. One can directly see that adding terms in the $z$-series has a weaker effect on the approximated $d_\mathrm{L}(z)$ function than in the $f$-one. Hence, even though the $z$-series seems at first more likely to invalidate the Taylor approximation, it turns out to give better results than the $f$-ones.

Consequently, having decided which kind of series to use in our calibration, its order is left to be selected. As can be seen in Table \ref{tab:cal-chi-wodisp}, limiting the $z$-series to the third order  gives optimal values of both the information criteria, AIC and BIC. Furthermore, in Table \ref{tab:cal-param-wodisp}, the $d_{3,z}$ and $d_{4,z}$ Taylor parameters associated  to the fourth and fifth order series are not significantly different from zero. 

\begin{figure}
\includegraphics[width=\columnwidth]{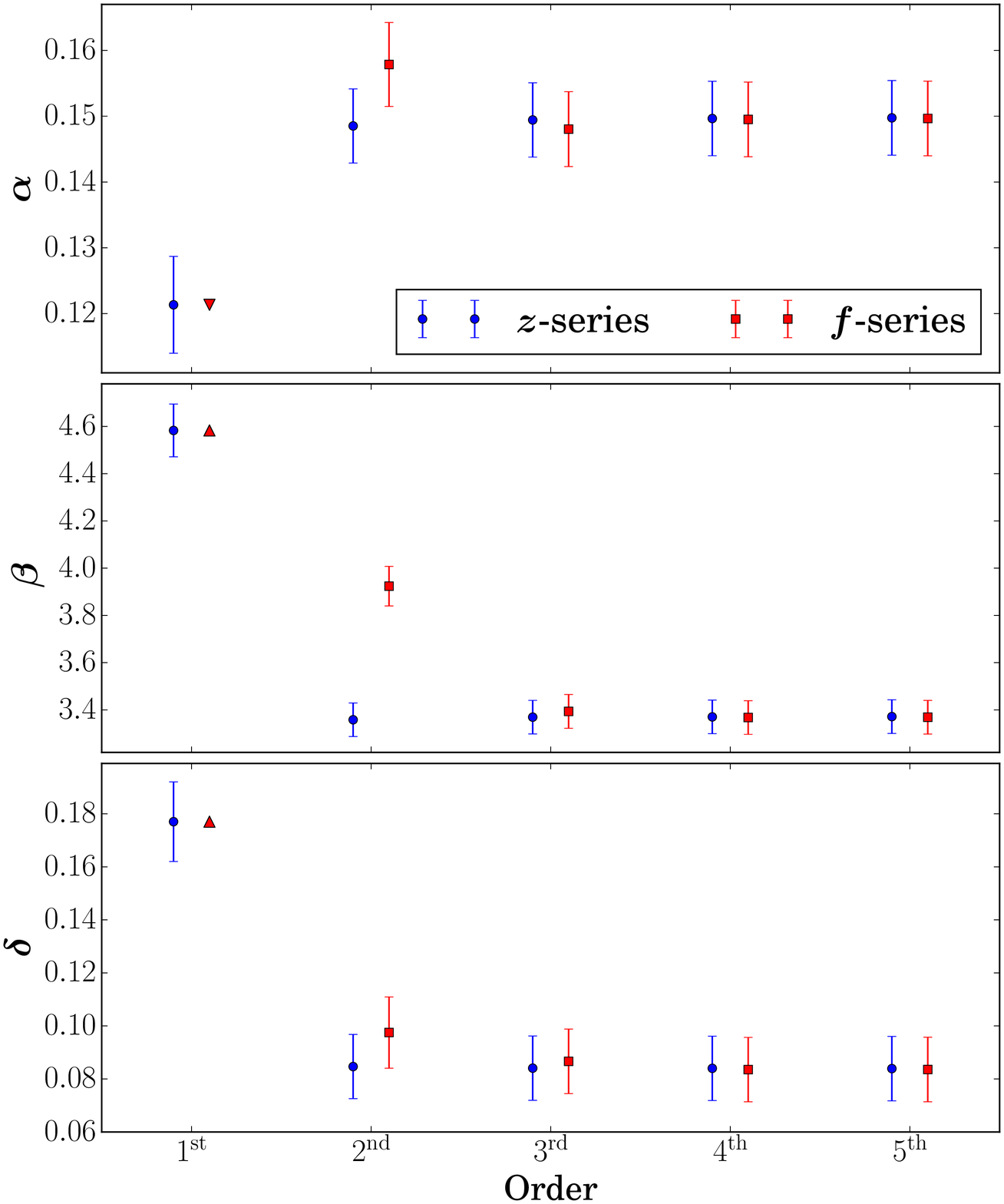}
\caption{Evolution of the standardisation parameters (\textit{from top to bottom}: $\alpha$, $\beta$ and $\delta$) as a function of the order of the limited $z$- (blue circles) and $f$-series (red squares). Exact values can be found in Table \ref{tab:cal-param-wodisp}. For purposes of readability, the parameters of the $f$-series limited to the first order are not depicted on the panels. Red triangles indicate their out-of-graph location. One can notice the non-significant variation of the parameters values from the second order of the $z$-series and from the third order of the $f$-ones. }
\label{fig:evol_sparam}
\end{figure}

On top of that, one can notice, when comparing lines of Table \ref{tab:cal-param-wodisp} or when observing the (unmoving) binned data points on Fig. \ref{fig:hub_diag_wodisp}, that the standardisation parameters $\alpha$, $\beta$ and $\delta$ do not significantly vary starting from the second order of the $z$-series or from the third one of the $f$-series. This is also illustrated on Fig. \ref{fig:evol_sparam}, showing the evolution of these parameters with the order of the Taylor series. Their values quickly stabilise for both the series. We could then even use the standardisation parameters values given by the fit of the second order of the $z$-series. One may notice that even if our calibration had suffered from overtraining, the parameters we are interested in (\textit{i.e.} the standardisation parameters) would not have been impacted by this issue.

For purposes of rigour, our cosmology-independent standardisation parameters will be chosen as the ones found through our $z$-series limited to the third order (highlighted in bold red in our Tables \ref{tab:cal-param-wodisp} and \ref{tab:cal-chi-wodisp}). The covariances are given in the matrix presented in Table \ref{tab:cal-covmat}. The SNe Ia data can then be properly recalibrated and used to perform fits of various cosmological models. \cite{Xu16}, for example, developed such a work on the raw JLA data. We want to highlight the fact that if using the same SNe Ia dataset to perform the calibration and the cosmological fits, the error propagation between the parameters of these two fits is quite tricky. We thus strongly encourage the reader to separate the SNe Ia at their disposal into two independent subsets, performing the calibration fit on one group of objects and the cosmological one on the other.

\begin{landscape}
\begin{table}
\caption{Best-fitting standardisation and Taylor parameters for our cosmology-independent methodology, up to different orders of both $z$ and $f$ Taylor series. In bold and red, one can find the calibration minimising the number of Taylor parameters while optimising the data representation (see Table \ref{tab:cal-chi-wodisp} associated). }
\label{tab:cal-param-wodisp}
\centering
\begin{tabular}{ c c c c c c c c c c }
\hline 
\hline
 & Order & $\alpha$ & $\beta$ & $\delta$ & $M_B+ 5 \log(d_{0, z/f})-5$ & $d_{1, z/f}$ & $d_{2, z/f}$ & $d_{3, z/f}$ & $d_{4, z/f}$ \\
\hline
\textbf{\boldmath$z$-series} & $1^\mathrm{st}$ & $ 0.121 \pm 0.007$ & $ 4.58 \pm 0.11 $ & $ 0.177 \pm 0.015 $ & $ 24.300 \pm 0.008 $ & $-$ & $-$ & $-$ & $-$  \\
 & $2^\mathrm{nd}$ & $ 0.149 \pm 0.006 $ & $ 3.358 \pm 0.071 $ & $ 0.085 \pm 0.012 $ & $ 24.074 \pm 0.009 $ & $ 0.606 \pm 0.018 $ & $-$ & $-$ & $-$  \\
 & \textcolor{red}{$\mathbf{3^\mathrm{rd}}$} & \textcolor{red}{$\mathbf{0.149 \pm 0.006}$} & \textcolor{red}{$\mathbf{3.369 \pm 0.071}$} & \textcolor{red}{$ \mathbf{0.084 \pm 0.012}$} & \textcolor{red}{$\mathbf{24.049 \pm 0.012}$} & \textcolor{red}{$\mathbf{0.726 \pm 0.044}$} & \textcolor{red}{$\mathbf{-0.151 \pm 0.050}$} & $-$ & $-$  \\
 & $4^\mathrm{th}$ & $ 0.150 \pm 0.006 $ & $ 3.370 \pm 0.071 $ & $ 0.084 \pm 0.012 $ & $ 24.044 \pm 0.015 $ & $ 0.761 \pm 0.075 $ & $ -0.26 \pm 0.19 $ & $ 0.08 \pm 0.13 $ & $-$ \\
 & $5^\mathrm{th}$ & $ 0.150 \pm 0.006 $ & $ 3.372 \pm 0.071 $ & $ 0.084 \pm 0.012 $ & $ 24.041 \pm 0.018 $ & $ 0.79 \pm 0.13 $ & $ -0.41 \pm 0.55 $ & $ 0.33 \pm 0.84 $ & $ -0.12 \pm 0.39 $  \\
 \hline
\textbf{\boldmath$f$-series} & $1^\mathrm{st}$  & $ - 0.381 \pm 0.065 $ & $ 33.2 \pm 2.8 $ & $ 0.68 \pm 0.10 $ & $ 24.934 \pm 0.063 $& $-$ & $-$ & $-$ & $-$  \\
 & $2^\mathrm{nd}$ & $ 0.158 \pm 0.006 $ & $ 3.924 \pm 0.084 $  & $ 0.098 \pm 0.013 $ & $ 23.822 \pm 0.013 $ & $ 3.500 \pm 0.060 $ & $-$ & $-$ & $-$ \\
 & $3^\mathrm{rd}$ & $ 0.148 \pm 0.006 $ & $ 3.393 \pm 0.071 $  & $ 0.087 \pm 0.012$ & $ 24.093 \pm 0.016 $ & $ 1.118 \pm 0.089 $ & $ 5.49 \pm 0.22 $ & $-$ & $-$ \\
 & $4^\mathrm{th}$ & $ 0.150 \pm 0.006 $ & $ 3.368 \pm 0.071 $  & $ 0.084 \pm 0.012 $ & $ 24.036 \pm 0.021 $ & $ 1.96 \pm 0.22 $ & $ 0.3 \pm 1.3 $ & $ 8.7 \pm 2.0 $ & $-$ \\
  & $5^\mathrm{th}$ & $ 0.150 \pm 0.006 $ & $ 3.369 \pm 0.071 $ & $ 0.084 \pm 0.012 $ & $ 24.038 \pm 0.027 $ & $ 1.88 \pm 0.42 $ & $ 1.4 \pm 3.8 $ & $ 3 \pm 13 $ & $ 7 \pm 14$ \\
\hline
\end{tabular}
\end{table}

\begin{table}
\caption{$\chi^2$, $\chi^2$ per degree of freedom (dof) and information criteria (AIC and BIC) associated to the best-fitting parameters values of Table \ref{tab:cal-param-wodisp}. For each of these calibrations, the fits have been made on all the $N = 740$ SNe Ia. In bold and red, one can find the model minimising the number of Taylor parameters while optimising the data representation (minimum values of both AIC and BIC).}
\label{tab:cal-chi-wodisp}
\centering
\subfloat{
\begin{tabular}{ c c c c c c c c c }
\hline 
\hline
 & Order & $k$ & $\chi^2$ & $\chi^2$/dof & AIC & $\Delta$AIC & BIC & $\Delta$BIC  \\
\hline
\textbf{\boldmath$z$-series} & $1^\mathrm{st}$ & $4$ & $2555$ & $ 3.47$ & $2563$ & $1522$ & $2581$ & $1512$  \\
 & $2^\mathrm{nd}$ & $5$ & $1038$ & $1.41$  & $1048$ & $7$ & $1071$ & $2$  \\
 & \textcolor{red}{$\mathbf{3^\mathrm{rd}}$} & \textcolor{red}{$\mathbf{6}$} & \textcolor{red}{$\mathbf{1029}$} & \textcolor{red}{$\mathbf{1.40}$} & \textcolor{red}{$\mathbf{1041}$} & \textcolor{red}{$\mathbf{0}$} & \textcolor{red}{$\mathbf{1069}$} & \textcolor{red}{$\mathbf{0}$}  \\
 & $4^\mathrm{th}$ & $7$ & $1029$ & $1.40$ & $1043$ & $2$ & $1075$ & $6$  \\
 & $5^\mathrm{th}$ & $8$ & $1029$ & $1.41$ & $1045$ & $4$ & $1082$ & $13$  \\
 \hline
\end{tabular}}
\hspace{0.5cm}
\subfloat{
\begin{tabular}{ c c c c c c c c c }
\hline 
\hline
 & Order & $k$ & $\chi^2$ & $\chi^2$/dof & AIC & $\Delta$AIC & BIC & $\Delta$BIC  \\
\hline
\textbf{\boldmath$f$-series} & $1^\mathrm{st}$ & $4$ & $4385$ & $5.96$ & $4393$  & $3352$ & $4411$ & $3342$ \\
 & $2^\mathrm{nd}$ & $5$ & $1556$ & $2.21$ & $1566$ & $525$  & $1589$ & $520$ \\
 & $3^\mathrm{rd}$ & $6$ & $1055$ & $1.44$ & $1067$ & $26$ & $1095$ & $26$ \\
 & $4^\mathrm{th}$ & $7$ & $1031$ & $1.41$ & $1045$ & $4$ & $1077$ & $8$ \\
 & $5^\mathrm{th}$ & $8$ & $1030$ & $1.41$ & $1046$ & $5$ & $1083$ & $14$ \\
\hline
\end{tabular}}
\end{table}

\begin{table}
\caption{Best-fitting standardisation and cosmological parameters along with their associated $\chi^2$, $\chi^2$ per degree of freedom (dof), AIC and BIC for simultaneous fits of different basic cosmologies. The present Hubble constant is assumed to be equal to 70 km/s/Mpc. The $\Delta$AIC$_{\mathrm{CI}}$ (respectively $\Delta$BIC$_{\mathrm{CI}}$) are defined as the differences between the corresponding AIC (resp. BIC) associated to the studied cosmology and the AIC (resp. BIC) of the best cosmology-independent (CI) calibration (here, the one built on the $z$-series limited to the third order). These latter can be found in bold red in Table \ref{tab:cal-chi-wodisp}. }
\label{tab:cal-cosmo}
\centering
\begin{tabular}{c c c c c c c c c c c c c c c}
\hline 
\hline
Cosmology & $\alpha$ & $\beta$ & $\delta$ & $M_B$ & $\Omega_{m,0}$ & $\Omega_{\Lambda,0}$ & $w$ & $k$ & $\chi^2$ & $\chi^2$/dof & AIC & $\Delta$AIC$_{\mathrm{CI}}$ &  BIC & $\Delta$BIC$_{\mathrm{CI}}$ \\ 
\hline 
Flat $\Lambda$CDM & $0.150\pm0.006$ & $3.379\pm0.071$ & $0.084 \pm 0.012$ & $-19.130\pm0.008$ & $0.273\pm0.018$ & $-$ & $-$ & $5$ & $1030$ &$1.40$ & $1040$ & $-1$ & $1063$ & $-6$ \\
$\Lambda$CDM & $0.149\pm0.006$ & $3.361\pm0.071$ & $0.084\pm0.012$ & $-19.108\pm0.013$ & $0.095\pm0.095$ & $0.44\pm0.14$ & $-$ & $6$ & $1030$ &$1.40$ & $1042$ & $+1$ & $1070$ & $+1$ \\
Flat $w$CDM & $0.149\pm0.005$ & $3.363\pm0.061$ & $0.084\pm0.012$ & $-19.108\pm0.012$ & $0.01^{+0.17}_{-0.01} $ & $-$ & $-0.60 \pm 0.15 $ & $6$ & $1030$ &$1.40$ & $1042$ & $+1$ & $1070$ & $+1$ \\
$w$CDM & $0.150\pm0.005$ & $3.371\pm0.061$ & $0.084\pm0.012$ & $-19.124\pm0.017$ & $0.15\pm0.21$ & $0.36\pm0.57$ & $-1.5\pm2.1$ & $7$ & $1029$ & $1.40$ & $1043$ & $+2$ & $1075$ & $+6$ \\
\hline
\end{tabular}
\end{table}

\end{landscape}

\begin{table}
\caption{Covariance matrix associated to our best cosmology-independent calibration (based on the $z$ Taylor series limited to the third order). The standardisation and Taylor parameters are presented in the same order than in Table \ref{tab:cal-param-wodisp}. }
\label{tab:cal-covmat}
\centering
\begin{tabular}{c}
\begin{equation*}
\textrm{Cov} = 10^{-6} \times
\begin{pmatrix}
32 & -60 & 21 & -13 & 22 & -16 \\
 & 5056 & 34 & -45 & 425 & -303 \\
 & & 147 & -46 & -32 & 9 \\
 & & & 147 & -436 & 410 \\
 & & & & 1929 & -1984 \\
 & & & & & 2457 \\  
\end{pmatrix}
\end{equation*}
\end{tabular}
\end{table}

\subsection{Digression about cosmology-dependent calibrations}
\label{Alt-dig}

For the sake of illustration, we also performed cosmology-dependent fits on SNe Ia data. These calibrations simultaneously fit the standardisation parameters as well as the cosmological ones as it was done by \cite{Suzuki12}, \cite{Betoule14} and many others since \cite{Perlmutter99}. These parameters as well as their associated $\chi^2$, AIC and BIC values are displayed in Table \ref{tab:cal-cosmo}, along with $\Delta$AIC$_{\mathrm{CI}}$ and $\Delta$BIC$_{\mathrm{CI}}$, the differences between the AIC/BIC values of the particular cosmology-dependent fit and those values for our best cosmology-independent (CI) calibration. From these values, one can notice that the cosmology-dependent calibrations have a goodness-of-fit similar to the best independent ones, the flat $\Lambda$CDM fit being marginally favoured by the data due to its smallest number of parameters.

With this article, we aim at testing the possible compliance of the SNe Ia data when fitting standardisation and cosmological parameters simultaneously. We thus develop a cosmology-independent calibration to compare results from dependent and independent methods. For the sake of this comparison, a clear distinction is made between those two families of calibrations: when focusing on finding the best independent one, we favour our Taylor expansion, even though the cosmology-dependent calibration results in slightly better fits (as it is the case here for the flat $\Lambda$CDM model).  On the other hand, when using SNe Ia data in general cosmological tests,  as done by \citet{Shi12} or \citet{Xu16} for example, and in case of a cosmology-dependent calibration leading to a much better fit than an independent one, we would advise to favour the former, despite the philosophical questions raised.

\subsection{Discussion}
\label{Alt-disc}

For readibility, the standardisation parameters that we found via our best cosmology-independent calibration as well as via a simultaneous fit with the flat $\Lambda$CDM model are summarized in Table \ref{tab:cal-param-wdisp}. Parameters from \cite{Betoule14} are also presented.

Comparing the first two lines of Table \ref{tab:cal-param-wdisp}, one can see that we retrieve quasi-identical parameters using either the simultaneous fit methodology or our cosmology-independent one. This means that even though assuming a particular cosmology to standardise the SNe Ia could theoretically lead to the data compliance problem, it is not the case in practice for the JLA compilation. JLA data are not significantly skewed towards the flat $\Lambda$CDM model assumed to process them. We thus independently confirm the results from \citet{Betoule14}, where the possible correlation between the cosmological and the standardisation parameters was found to be non-existent for this compilation. 

This can be explained by the fact that the numerous SNe Ia from the JLA compilation sufficiently populate the space of standardisation parameters and redshift. However, if that conclusion is valid for the present SNe Ia compilation, which includes hundreds of objects, it might not be valid for other standard candles, for which the sample might not be as large and as independent on redshift.  As an example, let us mention the superluminous supernovae (type Ic), for which only about 20 are presently known \citep{Inserra14, Inserra15, Wei15_Ic}. In such smaller samples, there is no guarantee that the space of standardisation parameters and redshift will be sufficiently filled so that the calibration will be independent of the assumed cosmological model. 


\begin{table*}
\caption{Best-fitting standardisation parameters for our cosmology-independent methodology (copied from the line highlighted in bold and red in Table \ref{tab:cal-param-wodisp}) and from a simultaneous fit technique similar to the one used by \citet{Betoule14} (copied from the first line of Table \ref{tab:cal-cosmo}). The $M_B$ parameter values are given assuming an present Hubble constant of $70$ km/s/Mpc. Our parameter values are also compared to  values from \citet{Betoule14} analysis. One will notice some significant differences.}
\label{tab:cal-param-wdisp}
\centering
\begin{tabular}{ c c c c c c  }
\hline 
\hline
& Methodology & $\alpha$ & $\beta$ & $\delta$ & $M_B$ \\ 
\hline
\multirow{2}{*}{Present work} & Cosmo-indpt & $ 0.149 \pm 0.006 $ & $ 3.369 \pm 0.077 $ & $ 0.084 \pm 0.012 $ & $-19.11 \pm 0.01$ \\ 
 & Sim. fit (flat $\Lambda$CDM) & $ 0.150 \pm 0.006 $ & $ 3.379 \pm 0.071 $ & $ 0.084 \pm 0.012 $ & $-19.13 \pm 0.01$ \\ 
\cite{Betoule14} & Sim. fit (flat $\Lambda$CDM) & $0.141 \pm 0.006$ & $3.101 \pm 0.075$ & $0.070 \pm 0.023$ & $-19.05 \pm 0.02$ \\ 
\hline
\end{tabular}
\end{table*}

\section{Conclusion}
\label{Ccl}

Even though the cosmology-dependency of SNe Ia data does not effectively appear in the JLA compilation, the philosophical interrogations raised at the end of Sec. \ref{Current} still remain open, valid and unsettled. In a sounder and more rigorous theoretical perspective, we would thus recommend the use of our simple, general and cosmology-independent methodology to determine the standardisation parameters, particularly because it can be applied to any standard candle sample. Our method could then be used in future works comparing cosmological model predictions to standard candle results, whatever the nature and size of the sample.

\section*{Acknowledgements}

Judith Biernaux acknowledges the support of the F.R.I.A. fund of the F.N.R.S.




\bibliographystyle{mnras}
\bibliography{Hauret_biblio}

\bsp	

\label{lastpage}
\end{document}